# Exact dynamics of quantum dissipative $XX$ models: Wannier-Stark localization in the fragmented operator space.

Alexander Teretenkov[1] and Oleg Lychkovskiy[2]

[1]*Department of Mathematical Methods for Quantum Technologies,*
*Steklov Mathematical Institute of Russian Academy of Sciences*
*8 Gubkina St., Moscow 119991, Russia*
[2]*Skolkovo Institute of Science and Technology*
*Bolshoy Boulevard 30, bld. 1, Moscow 121205, Russia*
(Dated: May 13, 2024)

We address dissipative dynamics of the one-dimensional nearest-neighbour $XX$ spin-1/2 chain governed by the Gorini-Kossakowski-Sudarshan-Lindblad (GKSL) equation. In the absence of dissipation the model is integrable. We identify a broad class of dissipative terms that generically destroy integrability but leave the operator space of the model fragmented into an extensive number of dynamically disjoint subspaces of varying dimensions. In sufficiently small subspaces the GKSL equation in the Heisenberg representation can be easily solved, sometimes in a closed analytical form. We provide an example of such an exact solution for a specific choice of dissipative terms. It is found that observables experience the Wannier-Stark localization in the corresponding operator subspace. As a result, the expectation values of the observables are linear combinations of essentially a few discrete decay modes, the long time dynamics being governed by the slowest mode. We examine the complex Liouvillian eigenvalue corresponding to this latter mode as a function of the dissipation strength. We find an exceptional point at a critical dissipation strength that separates oscillating and non-oscillating decay. We also describe a different type of dissipation that leads to a single decay mode in the whole operator subspace. Finally, we point out that our exact solutions of the GKSL equation entail exact solutions of the Schrödinger equation describing the quench dynamics in closed spin ladders dual to the dissipative spin chains.

*Introduction.* Explicit solutions of a quantum many-body problem are always welcome, since they enrich our understanding of the inherently complex many-body physics and often expose interesting phenomena with clarity and accuracy not accessible otherwise. Complementary to their conceptual importance, explicit solutions may have direct laboratory applications, thanks to the unceasing progress of experimental techniques and rapid rise of quantum technologies [1, 2].

In the present Letter we address the dynamics of open one-dimensional nearest-neighbour $XX$ spin-1/2 chains whose dynamics is described by the Gorini-Kossakowski-Sudarshan-Lindblad (GKSL) master equation [3–6]. We work in the Heisenberg representation, where the time evolution of an observable is embodied in the corresponding time-dependent Heisenberg operator. The latter obeys the Heisenberg version of the GKSL equation.

In a generic many-body system, coupled GKSL equations include an exponentially large hierarchy of operators and are expected to be too complex to be manageable without approximations. However, the complexity of the problem can be reduced if the space of operators is fragmented into sectors invariant under the GKSL evolution. Such *operator-space fragmentation* [7–9] is known to occur for open systems with quadratic bosonic or fermionic Hamiltonians and with Lindblad operators that are either linear [10–14], or quadratic and Hermitian [12, 15–21], or unitary with linear or quadratic generators [22] (see also [21, 23–25]), as well as for various open systems with zero or classical-like Hamiltonians and quantum dissipation [7, 17, 22, 26–30], or even with interacting Hamiltonians and fine-tuned dissipation [17].

Here we reveal a broad class of dissipative spin chains beyond the aforementioned ones that feature operator-space fragmentation. Further, we show that within this class the dynamics of a set of physically relevant few-body observables is confined to a small invariant operator subspace and can be easily (and often analytically) tracked.

We work out in detail two particular instances of dissipative $XX$ models. The first one features $\sigma^z$ dephasing and has been studied previously [18, 20, 31–36]. We highlight that this type of dissipation leads to a universal decay on top of the coherent dynamics, with a single decay exponent for all observables within the subspace.

In the second example the effect of dissipation turns out to be much more intricate. Heisenberg operators get localized in the operator subspace due to an effect similar to the Wannier-Stark localization of a particle in a constant electric field. As a result, a discrete sequence of decay modes appears, with only a few of them contributing to a particular observable. We study in detail the behavior of the slowest decay mode that governs the dynamics at long times. We discover a singularity (an exceptional point [37], to be more precise) in the corresponding Liouvillian eigenvalue as a function of dissipation strength, and establish its physical role. Finally, we discuss the duality between open spin chains and closed spin ladders, and the implications of our findings to the quench dynamics in spin ladders.

*GKSL equation.* GKSL equation in the Heisenberg rep-

resentation reads [5, Sec. 3.2.3]

$$\partial_t O_t = i[H, O_t] + \mathcal{D}^\dagger O_t, \qquad (1)$$

$$\mathcal{D}^\dagger O_t \equiv \gamma \sum_v \left( L_v^\dagger O_t L_v - \frac{1}{2}\{L_v^\dagger L_v, O_t\} \right), \qquad (2)$$

with the initial condition $O_{t=0} = O$. Here $O_t$ and $O$ are operators of the observable of interest in the Heisenberg and Schrödinger representations, respectively, $H$ is the Hamiltonian (in the Schrödinger representation),[1] $\mathcal{D}^\dagger$ is the ajoint dissipation superoperator (dissipator), $L_v$ are Lindblad operators[2] and $\gamma$ is a real positive constant. The expectation value $\langle O \rangle_t$ of the observable $O$ evolves in time according to $\langle O \rangle_t = \text{tr}\, \rho_0\, O_t$, where $\rho_0$ is the initial state of the open system. In the limit of vanishing dissipation, $\gamma = 0$, the GKSL equation (1) reduces to the Heisenberg equation.

*Onsager strings.* Throughout the paper we consider one-dimensional systems of spins 1/2. We start from defining a special type of operators that we refer to as Onsager strings.[3] An Onsager string $[\alpha\, \alpha']_j^{j+n}$ of length $n+1 \geq 2$ is a product of $n+1$ Pauli matrices on consecutive sites with two matrices $\sigma^{x,y}$ at the ends and $n-1$ matrices $\sigma^z$ in the middle:

$$[\alpha\alpha']_j^{j+n} \equiv \sigma_j^\alpha \sigma_{j+1}^z \ldots \sigma_{j+n-1}^z \sigma_{j+n}^{\alpha'}, \quad \alpha, \alpha' \in \{x, y\}. \qquad (3)$$

One additionally defines Onsager strings of length 1 that are simply $\sigma_j^z$.

Onsager strings have recurrently emerged in studies of the $XX$ and related models [38–41]. A standard way to deal with these models is to map spins to fermions through the Jordan-Wigner transformation [39]. In the fermionic representation, Onsager strings are nothing else but quadratic operators. It is useful to keep in mind this fact; however, we will not use the fermionic representation, since it, in general, perplexes the discription of dynamics, as discussed in what follows.

The real linear subspace of operators spanned by Onsager strings (Onsager space, for short) has the dimension $\sim 4N^2$, where $N$ is the number of spins, in contrast to the dimension $4^N$ of the whole operator space.

---

[1] Throughout the paper the presence (absence) of the subscript $t$ indicates that the operator is in the Heisenberg (Schrödinger) representation.
[2] The subscript $v$ in $L_v$ is somewhat schematic; specific way of enumeration of the Lindblad superoperators will be chosen on the case-by-case basis.
[3] There seems to be no universally accepted term for these operators. The rationale for the term chosen here is that these operators are building blocks for a representation of the Onsager algebra [38] (we do not use the latter, though).

*Onsager space invariance.* The key property of Onsager strings is that the Onsager space is closed with respect to commutation. This can be verified directly [41] or inferred from the fermionic representation.

As a consequence, the Onsager space is invariant under any Hamiltonian evolution generated by the Hamiltonian that itself belongs to the Onsager space. This class of Hamiltonians contains, in particular, paradigmatic $XX$, $XY$ and transverse-field Ising models.

Turning to the dissipative evolution (1), we enquire when the dissipator leaves the Onsager space invariant.

One can verify immediately that this is the case for self-adjoint Lindblad operators that belong themselves to the Onsager space (this can be most easily shown by using the equality $\mathcal{D}^\dagger O_t = \mathcal{D} O_t = -(\gamma/2)\sum_v [L_v, [L_v, O_t]]$ valid in the case of $L_v^\dagger = L_v$ ) as well as for unitary Lindblad operators with generators from the Onsager space. These facts, usually presented in the fermionic picture, are well-known [12, 15–22].

It turns out that, remarkably, there are options other than the above two cases. In fact, Lindblad operators need not be built from Onsager strings to keep the Onsager space invariant. Consider, for example, a Lindblad operator equal to $\sigma_j^x$. It does not belong to the Onsager space, yet it is easy to verify that it leaves the Onsager space invariant, since the corresponding term in the dissipator simply multiplies any Onsager string containing $\sigma_j^z$ or $\sigma_j^y$ by (-2) and annihilates other Onsager strings.

More complex Lindblad operators can be built by combining Onsager strings and operators $\sigma_j^{x,y}$. We summarize proper combinations in the following

**Lemma.** The evolution generated by the GKSL equation (1) leaves the Onsager space invariant provided that

(a) the Hamiltonian belongs to the Onsager space and

(b) each Lindblad operator

  (i) belongs to the Onsager space, or
  
  (ii) is a unitary operator with a generator from the Onsager space, or
  
  (iii) has the form
  
  $$p_x\, \sigma_j^x + p_y\, \sigma_j^y, \quad p_x, p_y \in \mathbb{R}, \quad \text{or} \qquad (4)$$
  
  (iv) is a product of any number of operators of the form (ii) and (iii).

A remark on the merit of the fermionic representation is in order here. As already noted, in fermionic picture Lindblad operators of types (i) and (ii) correspond to well-studied cases of Lindblad operators that are quadratic [15] or unitary with quadratic generators [22], respectively. Importantly, the size of support of these operators coincides in both spin and fermionic representation. In contrast, operators of type (iii) acquire an extensively large support (i.e. become highly nonlocal) in the fermionic representation. Furthermore, they are odd in fermion operators and thus do not conserve the fermion number. For these reasons the fermionic picture is hardly suitable for describing the dissipative dynamics of type (iii).

*Operator space fragmentation.* Consider a subspace spanned by symmetrised products of two Onsager strings. Thanks to the identity $[a, \{b, c\}] = \{[a, b], c\} + \{b, [a, c]\}$, this subspace is left invariant by commutation with an Onsager string or a linear combination thereof. It can be verified directly that, more generally, this subspace remains invariant with respect to the quantum-dissipative evolution (1) under the conditions of the above Lemma. One can further consider subspaces spanned by products of larger numbers of Onsager strings. In fact, a half of the whole operator space gets fragmented in this way (while the other half contains operators that cannot be represented as products of Onsager strings). In the present Letter we focus on the dynamics within the lowest subspace in this tower (dubbed "Onsager space" here), leaving higher subspaces for further work.

*XX model: Hamiltonian dynamics.* As a specific example of a Hamiltonian from the Onsager space, we consider the Hamiltonian of the translation-invariant one-dimensional nearest-neighbour $XX$ spins-1/2 model,

$$H = \frac{1}{2} \sum_{j=1}^{N} \left( \sigma_j^x \sigma_{j+1}^x + \sigma_j^y \sigma_{j+1}^y \right), \qquad (5)$$

where subscripts $j$ and $j + N$ refer to the same site [39].

To set the stage for the analysis of the dissipative dynamics, we first address the Hamiltonian dynamics, i.e. dynamics without dissipation. To this end we solve coupled Heisenberg equations in the Onsager space (*cf.* [18, 41–43]).

For simplicity, we focus on the translation-invariant sector of the Onsager space. It is spanned by the following operators:

$$A^n = \sum_{j=1}^{N} [xx]_j^{j+n}, \quad A^{-n} = \sum_{j=1}^{N} [yy]_j^{j+n},$$
$$B^n = \sum_{j=1}^{N} [xy]_j^{j+n}, \quad B^{-n} = -\sum_{j=1}^{N} [yx]_j^{j+n}, \quad (6)$$

where $n \geq 1$, and $A^0 = -\sum_{j=1}^{N} \sigma_j^z$.

It is easy to verify that operators $H^n = (1/2)(A^n + A^{-n})$ and $Q^n = (1/2)(B^n + B^{-n})$ are Hamiltonian integrals of motion (HIoM), i.e. commute with the Hamiltonian, $[H, H^n] = [H, Q^n] = 0$ [38, 40, 41].

It turns to be convenient to introduce non-Hermitian operators

$$R^{\pm n} = (1/2)(A^n - A^{-n}) \pm (i/2)(B^n - B^{-n}), \quad n \geq 1. \quad (7)$$

Coupled Heisenberg equations for $R_t^n$ acquire a particularly simple form [44]:

$$\partial_t R_t^n = -2i \left( R_t^{n-1} + R_t^{n+1} \right), \quad n \geq 1, \qquad (8)$$

where $R_t^0$ is identically zero.

Solving a linear system of differential equations essentially reduces to diagonalizing its matrix (if the latter is diagonalizable). The matrix of eq. (8) is very simple – its eigenvectors are plane waves. Standard calculations analogous to those in [43] (see the Supplement [44] for details) lead to

$$R_t^n|_{\gamma=0} = \sum_{m=1}^{\infty} i^{n-m} \Big( J_{m-n}(4t) - (-1)^n J_{m+n}(4t) \Big) R^m, \qquad (9)$$

where $J_{n\pm m}(4t)$ are Bessel functions. For further purposes, we explicitly indicate in the above formula that the dissipation is absent. The Heisenberg operators $A_t^n$, $B_t^n$ can be obtained from eq. (9), see [44].

To illustrate real-time quench dynamics, we consider a translation-invariant out-of-equilibrium initial state

$$|\text{in}\rangle = |\text{xxx}\ldots\text{x}\rangle, \qquad (10)$$

where all spins are polarized along the $x$ direction, and a simple observable $\sigma_j^x \sigma_{j+1}^x$. Eq. (9) entails [44]

$$\langle \sigma_j^x \sigma_{j+1}^x \rangle_t|_{\gamma=0} = (1/2)\big(1 + J_0(4t) + J_2(4t)\big). \qquad (11)$$

*XX model: $\sigma^z$ dissipation.* As a first example of a dissipative model, we consider the $XX$ model with Lindblad operators given by

$$L_j = \sigma_j^z, \qquad j = 1, 2, \ldots, N. \qquad (12)$$

These Lindblad operators satisfy conditions (i) and (ii) of the Lemma. This dissipative model and closely related ones have been extensively studied previously [18, 20, 31–36]. The model can be mapped to a fermionic model with a quadratic Hamiltonian and quadratic and Hermitian Lindblad operators [31]. A nonequilibrium steady state has been found in the case of a non-translation-invariant chain with biased boundaries [31, 32]. The GKSL equation in the Heisenberg representation has been solved in [18, 20, 35]. The model has been mapped to non-Hermitian Hubbard model in [33] (see also [18, 34]).

We reconsider this model within our framework. A remarkable feature of this model is that all operators (6) are eigen operators of the dissipator with the *same* eigenvalue:

$$\mathcal{D}F^{\pm n} = -4\gamma F^{\pm n}, \quad n \geq 1, \qquad (13)$$

where $F$ stands for $A$, $B$, $H$, $Q$ or $R$. The only exception from this rule is $A^0 = H^0$ that satisfies $\mathcal{D}H^0 = 0$ and thus remains a conserved quantity, in contrast to other HIoMs. Eq. (13) implies that the matrix of the corresponding coupled GKSL equations acquires a dissipative term of the form $-4\gamma \cdot \mathbb{1}$, where $\mathbb{1}$ is the identity matrix. As a consequence, all Heisenberg operators within the translation-invariant sector of the Onsager space (apart from $A^0$) acquire a universal damping exponent $e^{-4\gamma t}$ on top of the coherent dynamics,[4]

$$F_t^{\pm n} = e^{-4\gamma t}\left(F_t^{\pm n}|_{\gamma=0}\right), \quad n \geq 1. \quad (14)$$

Since the Hamiltonian dynamics, $F_t^n|_{\gamma=0}$, has been already found, we immediately obtain the dynamics in the presence of $\sigma^z$-dissipation (see [44] for an illustration).

*XX model: $\sigma^{x,y}$ dissipation.* Now we turn to a different type of dissipator:[5]

$$L_{2j-1} = (1/\sqrt{2})\,\sigma_j^x, \quad L_{2j} = (1/\sqrt{2})\,\sigma_j^y, \quad j = 1, 2, \ldots, N. \quad (15)$$

It satisfies condition (iii) of the Lemma.

In contrast to the previous case, the eigenvalue of the dissipator grows with the support of the eigen operator:

$$\mathcal{D}A^0 = -2\,\gamma\,A^0, \quad \mathcal{D}F^{\pm n} = -2\,\gamma\,n\,F^{\pm n}, \quad n \geq 1. \quad (16)$$

This expression immediately implies the support-dependent damping of HIoMs (see also [47]),

$$H_t^n = e^{-2\gamma n t}H^n, \quad Q_t^n = e^{-2n\gamma t}Q^n, \quad n \geq 1, \quad (17)$$

and $H_t^0 = e^{-2\gamma t}H^0$.

The dynamics of observables other than HIoMs is more complex. To see this, we again focus on $R_t^n$. The corresponding GKSL equations read

$$\partial_t R_t^n = -2i\left(R_t^{n-1} + R_t^{n+1}\right) - 2\,\gamma\,n\,R_t^n, \quad n \geq 1. \quad (18)$$

If $\gamma$ were imaginary, these equations would describe a quantum particle hopping on a half-line in a constant electric field; it is well-known that such a particle experiences Wannier-Stark localization [48]. Remarkably, it turns out that the localization phenomenon remains when the value of the "electric field" is imaginary. This can be seen by examining the eigenvectors of the matrix of eq. (18). The $l$'th eigenvector $\mathcal{U}_l^n$ and eigenvalue $\lambda_l$ read [44, 49]

$$\mathcal{U}_l^n = c_l J_{\nu_l+n}\left(-\frac{2i}{\gamma}\right), \quad \lambda_l = 2\,\gamma\,\nu_l, \quad l, n = 1, 2, \ldots, \quad (19)$$

where $c_l$ are normalization factors given in the Supplement [44] and $\nu_l$ are solutions of the equation

$$J_{\nu_l}(-2i/\gamma) = 0, \quad (20)$$

ordered by the descending real part. It should be stressed that the localization emerges for any nonzero dissipation strength.

---

[4] We remark that a similar exponential damping on top of a coherent dynamics has been found theoretically [45, 46] and experimentally [46] in finite $XXZ$ spin chains with dissipation.
[5] In the present case the choice of Lindblad operators is not unique: an equivalent choice that leads to the same dissipator reads $L_{2j-1} = \sigma_j^-$, $L_{2j} = \sigma_j^+$ (see e.g. [6, Chapt. 2]).

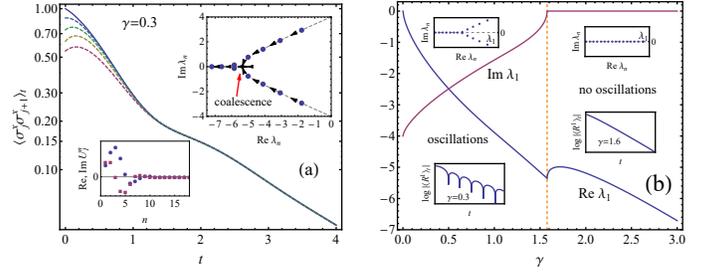

FIG. 1. Dynamics of dissipative $XX$ models. (a) Expectation value of $\langle\sigma_j^x\sigma_{j+1}^x\rangle_t$ after a quench from the initial state (10) for the $\sigma^{x,y}$-dissipation (15) (solid) and its approximations by 1, 2, 3 and 4 discrete modes (dashed, from bottom to top). Approximation by 5 modes is already indistinguishable from the exact result. Left inset: Wannier-Stark localization of an eigenvector of the matrix of eq. (18). Shown are real (blue circles) and imaginary (magenta squares) parts of the first eigenvector $\mathcal{U}_1^n$. Right inset: Liouvillian spectrum within the Onsager subspace. When varying the dissipation strength $\gamma$, eigenvalues closely follow a common trajectory in the complex plane shown by dashed line (arrows indicate the spectral flow direction when increasing $\gamma$, see the supplemental animation [44]). (b) Real and imaginary parts of the eigenvalue $\lambda_1$ as a function of the dissipation strength $\gamma$ in the case of $\sigma^{x,y}$-dissipation. This eigenvalue determines the leading dissipation mode at large times. Dashed orange line marks the critical value $\gamma_c \simeq \pi/2$. For $\gamma < \gamma_c$ the spectrum features at least one pair of complex roots leading to oscillating decay modes. For $\gamma > \gamma_c$ all Lindbladian eigenvalues are real, leading to pure decay without oscillations.

Thus obtained spectrum is shown in Fig. 1(a), its evolution with $\gamma$ is shown in the supplementary animation [44]. The spectrum has the following features (see [44, 49–53] for details). There is a phase transition at a critical dissipation strength $\gamma_c \simeq \pi/2$. If $\gamma \geq \gamma_c$, all eigenvalues are real, otherwise there is $n_p$ conjugate pairs of complex eigenvalues. $n_p$ is well approximated by the integer part of the ratio $\gamma_c/\gamma$. For $l \gtrsim 2n_p+2$, the eigenvalues are real and well approximated by $\lambda_l \simeq -2\gamma\,l$.

If one varies $\gamma$, eigenvalues move on the complex plain closely following a common trajectory. There is a discrete sequence of exceptional points $\gamma_1 = \gamma_c$, $\gamma_n \simeq \pi/(2n)$, where pairs of complex conjugate eigenvalues coalesce and the Lindbladian becomes non-diagonalizable. Exceptional points in open systems are known to have rich phenomenology and applications [37, 54–56], but we leave their detailed analysis for further work.

The eigenvectors (19) are exponentially localized in the vicinity of $n \simeq l$, see Fig. 1(a). Following a standard procedure [44], we obtain

$$R_t^n = \sum_{l,m=1}^{\infty} e^{\lambda_l t}\mathcal{U}_l^n\mathcal{U}_l^m\,R^m. \quad (21)$$

The localization implies that, in fact, a finite, independent of the system size number of terms in the above sum suffice to approximate $R_t^n$. As a result, an observable is well approximated by a few discrete decay modes, as illustrated in Fig. 1(a). This is the major consequence of localization.

Since the spectrum (19) is discrete, a single mode $\mathrm{Re}\,e^{\lambda_1 t}$ dominates the dynamics in the large time limit. The oscillatory part of this mode vanishes above the critical dissipation strength, as shown in Fig. 1 (b).

Localization in the *Krylov space* of operators explored by an observable in the Heisenberg representation has been recently discussed in the context of the recursion method and the growth of Krylov complexity in generic open systems [57, 58]. It is therefore plausible that the localization reported here is an exactly solvable example of a fairly generic phenomenon.[6]

*Duality between open and closed systems. (Non)integrability.* An open system can be mapped to a formally closed system with doubled degrees of freedom and a non-Hermitian Hamiltonian (see e.g. [59]). If the latter Hamiltonian is integrable (e.g. by Bethe ansatz), the corresponding open system can also be regarded as integrable [18, 33, 34, 60–62].[7] Dissipative spin chains studied here map to closed spin ladders [18, 33, 34, 44]. Importantly, models satisfying conditions (iii) or (iv) are generically nonintegrable in the above sense, the model (15) being no exception [70–73]. Instead, the corresponding closed spin ladders feature Hilbert space fragmentation [70, 74–76] – a closed system analog of operator space fragmentation.

If Lindblad operators are Hermitian, the Hamiltonian of the corresponding closed system can be made Hermitian by replacing $\gamma \to i\gamma$ [44]. Solutions of the GKSL equation then map to solutions of the Schrödinger equation for the dual closed system. This way we obtain exact solutions for the quench dynamics in the spin-ladder models dual to dissipative $XX$ models [44].

*Outlook.* Much effort is being invested in the studies of nonequilibrium steady states (NESS) in systems with biased boundary dissipation [15, 16, 31, 32, 77–82]. It would be interesting to apply our approach in such a setting. This will necessitate considering dissipators that describe gain and loss. It also seems promising to extend our approach to dissipative versions of models where coupled Heisenberg equations have been explicitly solved, such as chiral Potts models [43, 83] and Kitaev honeycomb model [84, 85].

---

[6] We also note that a simpler version of the operator localization has been found earlier in a spin systems with a classical Hamiltonian and diagonal-preserving dissipator [26].

[7] We accept this as a working definition of integrable open systems. For other approaches see [63–67]. Note that a rigorous definition of quantum integrability is a matter of an ongoing debate [68, 69].

*Acknowledgments.* We thank V. Gritsev for useful discussions and I. Ermakov for the careful reading of the manuscript. We are grateful to V.I. Yudson and V. Popkov for insightful remarks. The work of O. Lychkovskiy was supported by the Foundation for the Advancement of Theoretical Physics and Mathematics "BASIS" under the grant N⁰ 22-1-2-55-1. The work of A. Teretenkov was performed at the Steklov International Mathematical Center and supported by the Ministry of Science and Higher Education of the Russian Federation (agreement no. 075-15-2022-265).

# SUPPLEMENTARY MATERIAL
## TO THE LETTER
### "EXACT DYNAMICS OF QUANTUM DISSIPATIVE $XX$ MODELS: WANNIER-STARK LOCALIZATION IN THE FRAGMENTED OPERATOR SPACE" BY ALEXANDER TERETENKOV AND OLEG LYCHKOVSKIY.

## S1. DYNAMICS OF DISSIPATIVE $XX$ MODELS

### Systems of linear differential equations with symmetric matrices

Here we remind some basic facts from the theory of linear differential equations. We write down the coupled equations of motion for operators $R^n$ in the matrix form

$$\partial_t R_t = \mathcal{M} R_t, \qquad R_0 = R. \tag{S1}$$

Here $R_t = \|R_t^n\|_{n=1,2,\ldots}$ is a column vector constructed of Heisenberg operators $R_t^n$, $R = \|R^n\|_{n=1,2,\ldots}$ is a column vector constructed of Schrödinger operators $R^n$, and $\mathcal{M} = \|M_m^n\|_{m,n=1,2,\ldots}$ is some matrix.

We seek to express the Heisenberg operators $R_t^n$ as linear combinations of Schrödinger operators $R_t^n$. In matrix form this can be written as

$$R_t = \mathcal{G}(t) R, \qquad \mathcal{G}(0) = \mathbb{1}, \tag{S2}$$

where the time-dependent matrix $\mathcal{G}(t)$ is referred to as propagator. Then coupled differential equations (S1) are equivalent to

$$\partial_t \mathcal{G}(t) = \mathcal{M} \mathcal{G}(t), \qquad \mathcal{G}(0) = \mathbb{1}. \tag{S3}$$

The formal solution of this equation reads

$$\mathcal{G}(t) = e^{t\mathcal{M}}. \tag{S4}$$

We will deal with matrices $\mathcal{M}$ that are diagonalizable,

$$\mathcal{M}\mathcal{U} = \mathcal{U}\Lambda, \tag{S5}$$

where $\mathcal{U}$ is an invertible matrix with column vectors being eigenvectors of $\mathcal{M}$, and $\Lambda$ is the diagonal matrix with the corresponding eigenvalues on the diagonal. Then eq. (S4) can be written as

$$\mathcal{G}(t) = \mathcal{U} e^{\Lambda t} \mathcal{U}^{-1}. \tag{S6}$$

Furthermore, matrices $\mathcal{M}$ that we will encounter turn out to be symmetric:

$$\mathcal{M} = \mathcal{M}^\mathsf{T}. \tag{S7}$$

This implies that $\mathcal{U}$ can be chosen to satisfy $\mathcal{U}^{-1} = \mathcal{U}^\mathsf{T}$, which is equivalent to normalizing $\mathcal{U}$ according to

$$\mathcal{U}^\mathsf{T} \mathcal{U} = \mathbb{1}. \tag{S8}$$

Then eq. (S6) entails

$$\mathcal{G}(t) = \mathcal{U} e^{\Lambda t} \mathcal{U}^\mathsf{T} \mathcal{G}(0). \tag{S9}$$

Thus the problem of solving coupled equations (S1) is essentially reduced to diagonalizing $\mathcal{M}$.

### Hamiltonian dynamics

Here we show in more details how to solve Heisenberg equations for the integrable $XX$ model in the absence of dissipation. For convenience, we reiterate some definitions and formulae from the main text.

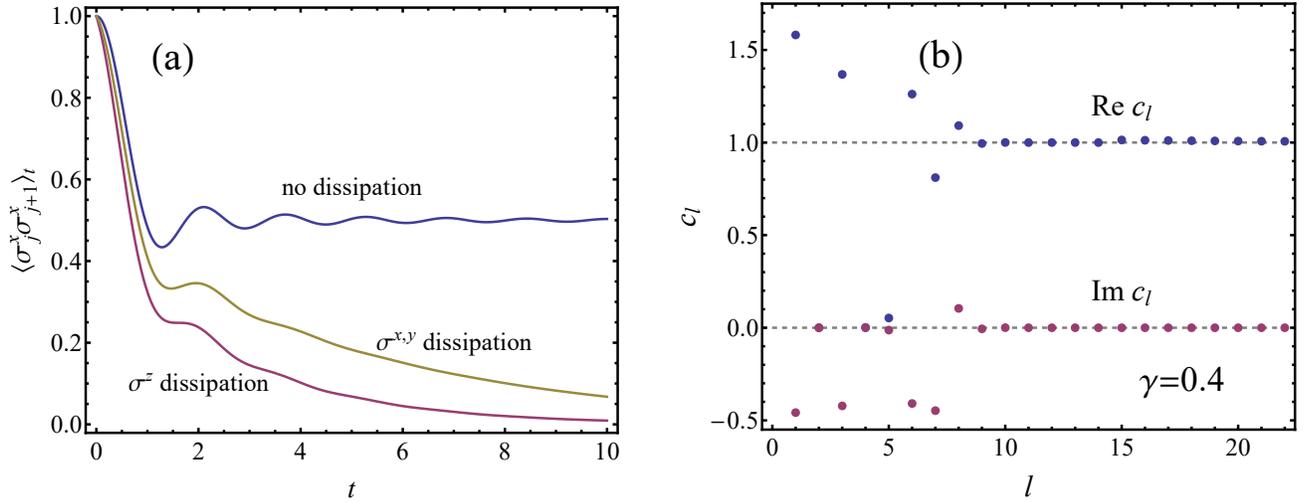

FIG. S1. (a) Expectation value of $\langle \sigma_j^x \sigma_{j+1}^x \rangle_t$ after a quench from the initial state (10) in three cases: no dissipation, $\sigma^z$-dissipation (12) and $\sigma^{x,y}$-dissipation (15). In the latter two cases $\gamma = 0.1$. (b) The normalization constant $c_l$ for $\gamma = 0.4$. One can see that for $l \gtrsim 2n_p + 2 \simeq 8$ the normalization constant is close to 1.

We introduce the following translation-invariant operators

$$A^n = \sum_{j=1}^{N} \sigma_j^x \left( \prod_{m=1}^{n-1} \sigma_{j+m}^z \right) \sigma_{j+n}^x, \qquad B^n = \sum_{j=1}^{N} \sigma_j^x \left( \prod_{m=1}^{n-1} \sigma_{j+m}^z \right) \sigma_{j+n}^y,$$

$$A^{-n} = \sum_{j=1}^{N} \sigma_j^y \left( \prod_{m=1}^{n-1} \sigma_{j+m}^z \right) \sigma_{j+n}^y, \qquad B^{-n} = -\sum_{j=1}^{N} \sigma_j^y \left( \prod_{m=1}^{n-1} \sigma_{j+m}^z \right) \sigma_{j+n}^x,$$

$$H^n = \frac{1}{2}(A^n + A^{-n}), \qquad Q^n = \frac{1}{2}(B^n + B^{-n}),$$

$$R^n = \frac{1}{2}(A^n - A^{-n}) + \frac{i}{2}(B^n - B^{-n}), \tag{S10}$$

where $n \geq 1$, and

$$A^0 = H^0 = -\sum_{j=1}^{N} \sigma_j^z, \qquad B^0 = Q^0 = R^0 = 0. \tag{S11}$$

Subscripts $j$ and $j+N$ refer to the same site, which ensures translation invariance. The Hamiltonian of the model is $H = H^1$. $H^n$ and $Q^n$ are Hamiltonian integrals of motion (HIoMs). $A^n$ and $B^n$ can be expressed through $R^n$ as

$$A^n = H^n + \frac{1}{2}\left(R^n + (R^n)^\dagger\right), \qquad B^n = Q^n - \frac{i}{2}\left(R^n - (R^n)^\dagger\right),$$

$$A^{-n} = H^n - \frac{1}{2}\left(R^n + (R^n)^\dagger\right), \qquad B^{-n} = Q^n + \frac{i}{2}\left(R^n - (R^n)^\dagger\right), \quad n \geq 0. \tag{S12}$$

Note that these relations remain valid if all Schrödinger operators are replaced by Heisenberg operators. We also note in passing that operators $A^n$ and $G^n = (i/2)(B^n - B^{-n})$ constitute a representation of the Onsager algebra (see e.g. [43]).

The matrix of coupled Heisenberg equations for the operators $R_t^n$, $n = 1, 2, \ldots$ reads

$$\mathcal{M} = -2 \begin{pmatrix} 0 & i & 0 & 0 & \ldots \\ i & 0 & i & 0 & \ldots \\ 0 & i & 0 & i & \ldots \\ 0 & 0 & i & 0 & \ldots \\ \vdots & \vdots & \vdots & \vdots & \ddots \end{pmatrix}, \tag{S13}$$

see eq.(8). This matrix describes the hopping of a quantum particle on a half-line. The $l$'th eigenvector and a corresponding eigenvalue of this matrix read

$$\mathcal{U}_l^n = c_l \sin(p_l \, n), \qquad \lambda_l = -4i \cos p_l, \qquad l, n = 1, 2, \ldots, \tag{S14}$$

where $p_l$ and $c_l$ are some real numbers. At this point, one can proceed in two equivalent ways. One way is to consider a spin chain of a finite size $N$ first, and then take the thermodynamic limit $N \to \infty$. In this case $p_l$ are quantized and admit $N$ distinct real values. The constants $c_l$ are then chosen to satisfy eq. (S8). This procedure has been used in ref. [43].

The second way, somewhat more elegant and concise, is to consider an infinite system from the outset. Then $p_l = p$ is an arbitrary real (to keep the eigenvector bounded) number from the interval $(-\pi, \pi]$, and the product of operators in eq. (S9) is understood as a convolution,

$$\mathcal{G}_m^n(t) = 2 \int_{-\pi}^{\pi} \frac{dp}{2\pi} e^{-4it \cos(p)} \sin(np) \sin(mp). \tag{S15}$$

The normalization here is chosen to ensure $\mathcal{G}_m^n(0) = \delta_m^n$. The integral above can be expressed through Bessel Functions, with the final result

$$\mathcal{G}_m^n(t) = i^{n-m} \Big( J_{m-n}(4t) - (-1)^n J_{m+n}(4t) \Big). \tag{S16}$$

Using eqs. (S2),(S12) one obtains explicit expressions for Heisenberg operators $R_t^n$ (see eq. (9) in the main text), $A_t^n$ and $B_t^n$. For example,

$$A_t^1 = H^1 + \sum_{k=1}^{\infty} (-1)^k \big( J_{2k-1}(4t) + J_{2k+1}(4t) \big) R^{2k-1}. \tag{S17}$$

The knowledge of a Heisenberg operator translates immediately into the knowledge of the dynamics of the corresponding observable. For the initial state (10),

$$|\text{in}\rangle = |\text{xxx} \ldots \text{x}\rangle, \tag{S18}$$

one obtains

$$\langle \text{in}|H^1|\text{in}\rangle = N/2 \qquad \langle \text{in}|R^{2k-1}|\text{in}\rangle = \delta_1^k \, N/2. \tag{S19}$$

Thus eq. (S17) leads to

$$\langle A^1 \rangle_t = N \big( 1 + J_0(4t) + J_2(4t) \big)/2. \tag{S20}$$

Due to translation invariance, for any $j$ one has $\langle \sigma_j^x \sigma_{j+1}^x \rangle_t = N \langle A^1 \rangle_t$, which leads to eq. (11) from the main text. The dynamics is illustrated in Fig. S1 (a).

### Dissipative dynamics with $\sigma^{x,y}$ dissipation

The matrix of coupled GKSL equations (18) reads

$$\mathcal{M} = -2 \begin{pmatrix} \gamma & i & 0 & 0 & \ldots \\ i & 2\gamma & i & 0 & \ldots \\ 0 & i & 3\gamma & i & \ldots \\ 0 & 0 & i & 4\gamma & \ldots \\ \vdots & \vdots & \vdots & \vdots & \ddots \end{pmatrix}, \tag{S21}$$

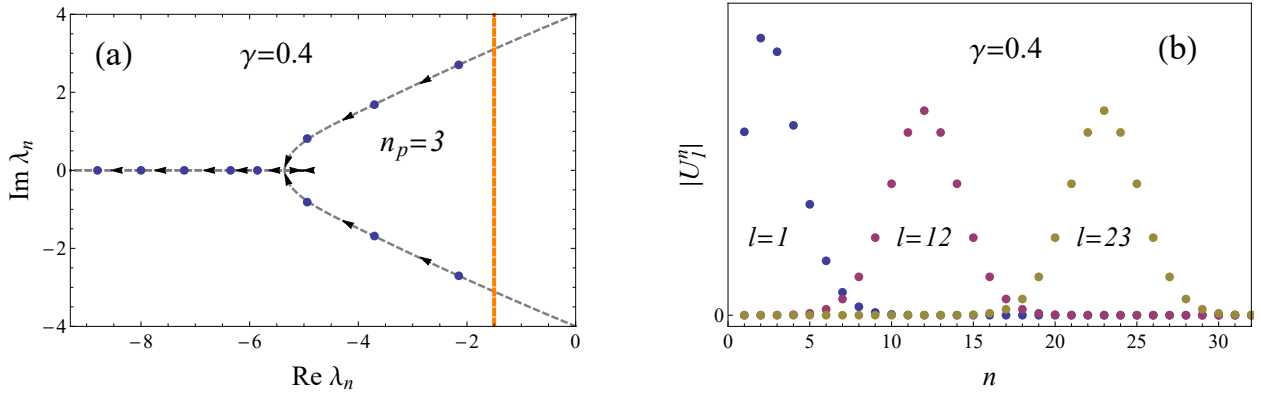

FIG. S2. (a) Spectrum of the matrix (S21) at $\gamma = 0.4$. There are $n_p = 3$ pairs of complex roots, other roots are real. Dashed orange vertical line corresponds to the upper bound (S26) from ref. [51]. Dashed gray line is the common trajectory closely followed by the eigenvalues when varying $\gamma$ (arrows indicate the direction of the spectral flow when increasing $\gamma$). (b) Wannier-Stark localization of the eigenvectors of the matrix of eq. (18). Shown are absolute values of the unnormalized first, 12'th and 23'd eigenvector. One can see that an $n$'th component of the $l$'th eigenvector vanishes outside the vicinity $n \simeq l$.

It describes hopping of a particle on a half-line with the imaginary "electric field" (for an analogous problem with a real electric field see e.g. [48]). The eigenproblem for this matrix has the same form as the well-known recurrence relations satisfied by Bessel functions, therefore the eigenvectors are expressed through the Bessel functions. Explicitly, the $l$'th eigenvector and eigenvalue read [49]

$$\mathcal{U}_l^n = c_l J_{\nu_l + n}\left(-\frac{2i}{\gamma}\right), \qquad \lambda_l = 2\gamma\nu_l, \qquad l, n = 1, 2, \ldots, \tag{S22}$$

where $\nu_l$ is a solution of the equation

$$J_{\nu_l}\left(-\frac{2i}{\gamma}\right) = 0. \tag{S23}$$

and $c_l$ is the normalization constant discussed below.

This way, solving the eigenproblem for the matrix (S21) reduces to solving the equation (S23). This equation has been studied in mathematical literature for a long time [49–53]. Here we give a description of its solution based on this prior work and on our numerical experiments.

The roots $\nu_l$ of eq. (S23) form a discrete sequence [49–51], see Fig. S2(a) (see also the inset to Fig 1(a) in the main text). The real parts of $\nu_l$ are negative [50], as expected on physical grounds.

We label $\nu_l$, $l = 1, 2, \ldots$ in the descending order of the real part, $\text{Re}\,\nu_{l+1} \leq \text{Re}\,\nu_l$, and if the real parts are equal, we choose $\text{Im}\,\nu_l < \text{Im}\,\nu_{l+1}$. For $\gamma$ below some critical value $\gamma_c$, first few roots are complex and split into complex-conjugate pairs, while other roots are real; for $\gamma \geq \gamma_c$ all roots are real [49]. As can be inferred from fig. 3.3 of ref. [49] and further verified numerically, the number of complex conjugate pairs $n_p$ can be approximated by

$$n_p \simeq [\gamma_c/\gamma], \tag{S24}$$

where [...] is the integer part. We numerically find

$$\gamma_c \simeq 1.5775. \tag{S25}$$

Rigorous bounds on the real and imaginary parts of $\nu_l$ are known [51, 52]:

$$\text{Re}\,\nu_l \leq \text{Re}\,\nu_1 \leq \begin{cases} -3/2, & \gamma < \gamma_c \quad [51], \\ -1, & \gamma \geq \gamma_c \quad [52], \end{cases} \tag{S26}$$

$$|\text{Im}\,\nu_l| < 2/\gamma \qquad [52]. \tag{S27}$$

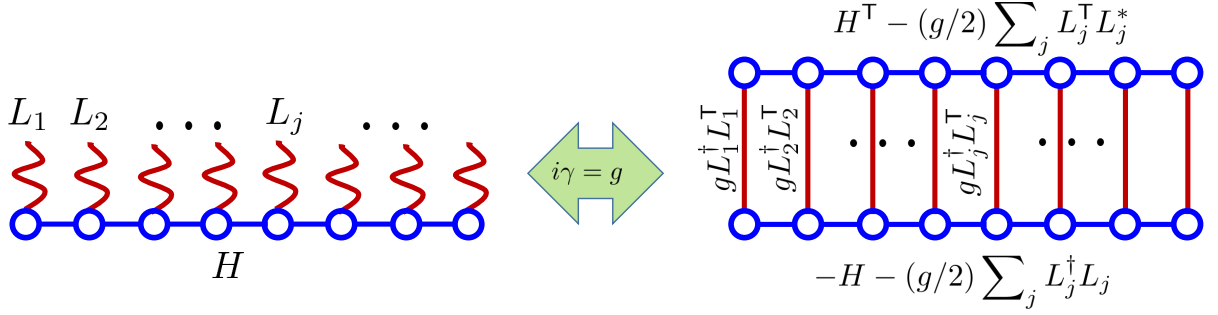

FIG. S3. An illustration of the duality (S34) between open systems and closed systems with doubled degrees of freedom.

We have verified numerically that starting from a certain $l$ the roots are well approximated by

$$\nu_l \simeq -l, \qquad l \gtrsim 2n_p + 2, \tag{S28}$$

see Fig. S2(a). This approximation has been rigorously justified for sufficiently large $\gamma$ [52]. It also follows from the following reasoning [48]. Consider an analog of the matrix (S21) that describes the hopping of the particle on a line instead of the half-line. The eigenproblem for such a matrix is well-known in the physics of Wannier-Stark localization, with the only difference that there $\gamma$ is imaginary. The eigenvector and eigenvalues of the latter problem are known to have the form (S22) with $\nu_l = l$ and $l$ running through all integers. Clearly, they will approximate the respective eigenvectors and eigenvalues of the original eigenproblem (S21) far from the origin, thanks to the localization.

The normalization constant $c_l$ is chosen to satisfy the normalization condition (S8). It can be found with the use of the identity [53]

$$J_\mu(z) J_{\nu+1}(z) - J_{\mu+1}(z) J_\nu(z) = 2\frac{\mu - \nu}{z} \sum_{m=1}^{\infty} J_{\mu+m}(z) J_{\nu+m}(z) \tag{S29}$$

in the limit $\mu \to \nu$. The result reads

$$c_l^{-2} = -\frac{i}{\gamma} J_{\nu+1}\left(-\frac{2i}{\gamma}\right) \partial_\nu J_\nu\left(-\frac{2i}{\gamma}\right)\bigg|_{\nu=\nu_l}. \tag{S30}$$

Since it is the square of $c_l$ that enters the physical observables (21), the sign of $c_l$ is not important.

We have verified numerically that $c_l^2 \to 1$ in the limit of $l \to \infty$. In fact, $c_l^2 \simeq 1$ already for $l \gtrsim 2n_p + 2$, as shown in Fig. S1(b). Consequently, the above-discussed approximate solution to the eigenproblem reads

$$\mathcal{U}_l^n \simeq J_{l-n}(2i/\gamma), \qquad \lambda_l \simeq -2\gamma l, \qquad l \gtrsim 2n_p + 2. \tag{S31}$$

## S2. DUALITY BETWEEN OPEN AND CLOSED SYSTEMS.

An adjoint Lindbladian,

$$\mathcal{L}^\dagger = i[H, \bullet] + \gamma \sum_v \left( L_v^\dagger \bullet L_v - \frac{1}{2}\{L_v^\dagger L_v, \bullet\} \right), \tag{S32}$$

is a linear superoperator acting in the space of operators. The duality between open systems and closed systems with doubled degrees of freedom is established by "vectorising" operators and treating the Lindbladian as an operator in the corresponding vector space [59]:

$$O = \sum_{m,n} O_{mn} |m\rangle\langle n| \quad \longleftrightarrow \quad |O\rangle = \sum_{m,n} O_{mn} |m\rangle \otimes |n\rangle, \tag{S33}$$

$$i\mathcal{L}^\dagger \longleftrightarrow \mathcal{H} = -H \otimes \mathbb{1} + \mathbb{1} \otimes H^\mathsf{T} + g \sum_v \left( L_v^\dagger \otimes L_v^\mathsf{T} - \frac{1}{2} L_v^\dagger L_v \otimes \mathbb{1} - \frac{1}{2} \mathbb{1} \otimes L_v^\mathsf{T} L_v^* \right), \tag{S34}$$

where $g = i\gamma$. Here one first fixes a basis $\{|n\rangle\}$ in the "original" Hilbert space (where operators of observables of the open system act) and then constructs a basis $\{|m\rangle \otimes |n\rangle\}$ in the doubled Hilbert space where the Lindbladian counterpart, $\mathcal{H}$, acts. The dual of the adjoint GKSL equation (1) then reads

$$i\partial_t |O\rangle_t = \mathcal{H} |O\rangle_t. \tag{S35}$$

As long as $\gamma$ is real, the coupling constant $g$ is imaginary, and $\mathcal{H}$ is non-Hermitian and does not describe a bona-fide closed system. However, our approach to solving GKSL equations is purely algebraic and therefore equally applies to imaginary $\gamma$ (i.e. real $g$). If $g$ is real and, in addition, Lindblad operators $L_v$ are Hermitian, $\mathcal{H}$ becomes Hermitian and describes a bona-fide closed system. Dissipative $XX$ models considered in the paper are dual to closed spin ladders with the $XX$ Hamiltonians acting along the legs and interactions of the form $L_v^\dagger \otimes L_v^\mathsf{T}$ acting along the rungs.

In particular, the model with the $\sigma^{x,y}$ dissipation (15) is dual to the $XX$ spin ladder with the Hamiltonian

$$\mathcal{H} = \frac{1}{2} \sum_j \left( -\sigma_j^x \sigma_{j+1}^x - \sigma_j^y \sigma_{j+1}^y + \tau_j^x \tau_{j+1}^x + \tau_j^y \tau_{j+1}^y + g \left( \sigma_j^x \tau_{j+1}^x + \sigma_j^y \tau_{j+1}^y \right) \right), \tag{S36}$$

up to an additive constant. Here $\sigma_j^\alpha$ and $\tau_j^\alpha$ are Pauli matrices acting on two different legs of the ladder.

Under the above duality, the operator space fragmentation maps to its closed system analogue – Hilbert space fragmentation (HSF) [74–76]. The Hilbert space fragmentation in a spin ladder similar to the ladder (S36) has been found in ref. [70].

Further, solutions of the GKSL equations (1) map to the solutions of the Schrödinger equation (S35) for the dual closed system. This means that the explicit solutions (14),(21) of the GKSL equations for the dissipative $XX$ models studied here immediately translate, with the help of the mapping (S33), to explicit solutions of the Schrödinger equation for the corresponding dual closed spin ladders. This way we get exact solutions for quantum quenches in closed spin ladders. Further elaboratoin of this research direction is left for future work.